\documentclass[a4paper,11pt]{amsart}
\begin{document}
\hyphenation{gra-vi-ta-tio-nal re-la-ti-vi-ty Gaus-sian
re-fe-ren-ce re-la-ti-ve gra-vi-ta-tion Schwarz-schild}
\title[On the displacements of Einsteinian fields \emph{et cetera}]
{{\bf  On the displacements of Einsteinian fields \\ \emph{et
cetera}}}
\author[Angelo Loinger]{Angelo Loinger}
%\date{}
%\address{Dipartimento di Fisica, Universit\`a di Milano, Via
%Celoria, 16 - 20133 Milano (Italy)}
%\email{angelo.loinger@mi.infn.it}
\thanks{To be
published on \emph{Spacetime \& Substance.} \\email:
angelo.loinger@mi.infn.it\\ Dipartimento di Fisica, Universit\`a
di Milano, Via Celoria, 16 - 20133 Milano (Italy)}

\begin{abstract}
I give here:  \emph{i}) a very simple proof that the physical
non-existence of gravitational waves (GW's) is quite consistent
with the basic principles of general relativity (GR); \emph{ii}) a
new argument against the physical existence of GW's; \emph{iii}) a
criticism of Fock's treatment of the GW's; \emph{iv}) some remarks
on recent experimental investigations concerning the GW's.
\end{abstract}

\maketitle

%%\begin{equation} \label{eq:sevenprime}
%%    \ddot{\Re} + \frac{\kappa}{6}\Re \rho=0 , \tag{7'}
%% \end{equation}
%% ``mechanisms'' \textrm{d} \`a
%% \cite{1}

 \vskip1.20cm
%\section{}
\noindent \textbf{1}. -- The following is a widespread and
erroneous opinion: \emph{Without gravitational waves (GW's), one
would have to explain an} \textbf{\emph{instantaneous}}
\emph{propagation of a change in the metric over the whole
universe simply by changing the distribution of stress or mass of
a given physical system}. -- In reality, the physical
non-existence of GW's is quite consistent with the principles of
general relativity (GR), as I have shown \emph{ad abundantiam} in
various papers \cite{1}, but perhaps in too concise ways insofar
as the above specific belief is concerned. I shall give now in
sects.\textbf{2}., \textbf{3}. a detailed treatment of it, with
the hope of convincing even the most naive among the physicists
that the adjective ``instantaneous'' is not the attribute of a
relativistic bugaboo -- if it is properly understood.

\par In sect.\textbf{4}. I give a new argument against the
physical existence of GW's. In sect.\textbf{5}. Fock's
computations concerning the GW's are critically examined. The
\textbf{\emph{Appendix}} reports some recent (negative) results of
the experimental search of GW's due to LIGO collaboration.

\vskip0.50cm
%\section{}
\noindent \textbf{2}. -- In previous Notes I have repeatedly
emphasized that Einstein field is \emph{not} analogous to Maxwell
field, since it has peculiar properties of its own that are not
shared by the electromagnetic field. If, however, we neglect for a
moment -- \emph{ad usum Delphini} -- the existence of the e.m.
waves, we can exploit a precise property of Maxwell field for our
purpose. For convenience, I utilize here the treatment of
Li\'enard-Wiechert e.m. fields -- created by a moving point charge
-- as is developed in the well known treatise by Becker and Sauter
\cite{2}; see in particular p.293 of this book, which gives the
expressions of the electric and magnetic fields due to Li\'enard
and Wiechert. For our aim, it is expedient to consider the first
part, say $\textbf{E}_{1}$, of the electric field $\textbf{E}$
(e.g.), i.e. the part which does not depend on the charge
acceleration. We have

\begin{equation} \label{eq:one}
    \textbf{E}_{1}(\tau)/e= \left[\frac{(\textbf{r}-r\textbf{v}/c)(1-v^{2}/c^{2})}
    {(r-\textbf{r}\cdot \textbf{v}/c)^{3}}\right]_{\tau \equiv t-r/c} \quad,
\end{equation}

with evident and standard notations. As Becker and Sauter write,
$\textbf{E}_{1}$ has the character of a \textbf{\emph{static}}
field, it falls off as $1/r^{2}$ for large distances. Since eq.
(\ref{eq:one}) gives the first part of the expression of the
global field $\textbf{E}$, which is valid for \emph{all}
velocities $\textbf{v}$, it must agree with the field, say
$\textbf{E}'$, created by a uniformly moving charged particle (see
sect.\textbf{64} of \cite{2}):

\begin{equation} \label{eq:two}
    \textbf{E}'(t)/e=
    \frac{\textbf{r}(1-v^{2}/c^{2})}{[r^{2}-(\textbf{r}\times\textbf{v}/c)^{2}]^{3/2}} \quad;
\end{equation}

the formal difference between expressions (\ref{eq:one}) and
(\ref{eq:two}) comes from the different meanings of the vector
$\textbf{r}$ in the two formulae. In eq.(\ref{eq:two})
$\textbf{r}=\textbf{r}(t)$ is set equal to the vector from the
\textbf{\emph{instantaneous}} particle location, say $B$, to the
field point $P$, while in eq.(\ref{eq:one}) by
$\textbf{r}=\textbf{r}(\tau)\equiv \textbf{r}(t-r/c)$ we
understand the vector from the particle location, say $A$, at
\textbf{\emph{retarded}} time $\tau \equiv t-r/c$, to the field
point $P$. For the case of \emph{constant} velocity we obviously
have:

\begin{equation} \label{eq:three}
    \textbf{r}(t) = \textbf{r}(\tau)-\frac{r(\tau)}{c}\textbf{v}
    \quad.
\end{equation}

If we write $\textbf{r}(t) \equiv \textbf{r}_{0}$, and
$\textbf{r}(\tau) \equiv \textbf{r}$ (as in eq.(\ref{eq:one})), we
find from $\textbf{r}_{0}=\textbf{r}-r\textbf{v}/c$ for the
denominator of eq. (\ref{eq:two}) that

\begin{equation} \label{eq:four}
    \left[r_{0}^{2}-\left(\textbf{r}_{0}\times
    \textbf{v}/c\right)^{2}\right]^{3/2} = \left(r-\textbf{r}\cdot
    \textbf{v}/c\right)^{3}    \quad,
\end{equation}

i.e. the denominator of eq. (\ref{eq:one}). Thus the field
$\textbf{E}_{1}$ actually represents the field \emph{moving along
with the particle}; and this is clearly true also for a
\emph{non}-constant speed. By contrast, the second part,
$\textbf{E}_{2}$, of the total electric field
$\textbf{E}=\textbf{E}_{1}+\textbf{E}_{2}$, which is proportional
to the acceleration \textbf{\.v}, has the character of a wavy
$(1/r)$ - decreasing field.

\par (I have reproduced almost literally some passages of Becker
and Sauter \cite{2}, only the italics are mine.)

\vskip0.50cm
%\section{}
\noindent \textbf{3}. -- We have seen that the \emph{static} part
$\textbf{E}_{1}$ of Li\'enard-Wiechert electric field $\textbf{E}$
\emph{moves} \textbf{\emph{en bloc}} \emph{with the particle}.
Now, the \textbf{\emph{same}} thing happens, in the
\textbf{\emph{exact}} formulation of GR, for the Einstein field
$g_{jk}(x^{0},\textbf{x})$ , $(j,k=0,1,2,3)$, since -- as it has
been proved \cite{3} -- \emph{no ``mechanism'' exists in GR, which
is capable of producing GW's}. In other terms, if we displace a
mass, its gravitational field and the related curvature of the
interested manifold \emph{displace themselves along with the
mass}. In general, qualitatively speaking, we can affirm that
under this respect Einstein field and Newton field behave in an
identical way. This fact is mathematically and physically
\emph{evident} in Friedmann's cosmological models, as I have shown
\cite{4}, owing to the perfect agreement between Friedmann's
solutions and the solutions of corresponding Newtonian models.
(Furthermore, we can remark that at any stage of the EIH-method of
solution of field equations there is a suitable reference frame
for which the solution has a \emph{Newtonian} form.)

\par Conclusion: the widespread opinion reported at the beginning
of sect.\textbf{1} is \emph{false}: the absence of GW's does not
generate any theoretical difficulty -- as Levi-Civita had pointed
out many years ago.

\par (Generally speaking, the real existence of \emph{physical}
waves requires the existence of \emph{physically} privileged
reference frames, or of a \emph{material} medium as the cosmic
ether. It is not the case of GR: in it a geodesic deviation must
have a \emph{Newton}-like character -- and therefore could be
recorded only by an apparatus in a relative proximity of the
gravity source.) --

\vskip0.50cm
%\section{}
\noindent \textbf{4}. - The Einstein field equations share with
Laplace-Poisson equation $\nabla^{2}U=-4 \pi G \rho$ an important
property. Let us consider for a moment only the case of a ``cloud
of dust'' with mass tensor $T^{jk}=\rho u^{j}u^{k}$,
$(j,k=0,1,2,3)$, where $\rho(x^{0},\textbf{x})$ is the invariant
mass density and $u^{j}(x^{0},\textbf{x})$ is the four-velocity of
a gravitating particle. It is well known \cite{5} that we can
always choose a Gaussian normal (``synchronous'' in Landau's
terminology \cite{6}) reference frame, for which:

\begin{equation} \label{eq:five}
\textrm{d}s^{2} = \left(\textrm{d}x^{0}\right)^{2} -
h_{\alpha\beta}\left(x^{0},\textbf{x}\right)
\textrm{d}x^{\alpha}\textrm{d}x^{\beta} \quad,
(\alpha,\beta=1,2,3) \quad;
\end{equation}

if there are only gravitational interactions -- as in the present
case --, this frame is also co-moving \cite{7}: the \emph{world}
lines of the ``dust'' particles are both \emph{time} lines and
\emph{geodesic} lines. Our mass tensor $T^{jk}$ has only the
component $T^{00}=\rho$ different from zero. \emph{Thus} --
exactly as it happens for Friedmann's models \cite{4} -- \emph{the
metric tensor} $g_{jk}(x^{0},\textbf{x})$ \emph{depends}
\textbf{\emph{only}} \emph{on} $\rho(x^{0},\textbf{x})$, in
perfect analogy with the Newtonian potential $U$, and it satisfies
\emph{identically} the geodesic equations. No GW's are emitted --
and this fact is now quite intuitive, because we see that the
motion of the fluid has been formally ``obliterated''.

\par This treatment can be immediately generalized to a continuum,
whose particles are subject to gravitational and
\emph{non}-gravitational (e.g., electromagnetic) interactions. It
is indeed sufficient to choose a \emph{co-moving} reference frame
-- as it is always possible if the particle trajectories do not
cross. Here too the metric tensor does not depend on the motion of
the medium -- motion that the metropolitan legend considers
responsible of the emission of GW's.

\vskip0.50cm
%\section{}
\noindent \textbf{5}. -- Fock \cite{8}
 pretended erroneously that the so-called \emph{harmonic frames}
 possess a \emph{physical} privilege with respect to the other
 co-ordinate systems. Thus, in particular, all his computations
 concerning the GW's are performed in a harmonic frame, and with
 mass tensors of \emph{extended} bodies. Since the motions of
 gravitating \emph{point} masses do not generate GW's, it is
 difficult to believe in a thaumaturgical virtue of largeness.
 Indeed, the extended bodies are composed of particles, and,
 further, their translational motions are correctly treated as
 motions of material corpuscles.

\par Fock's computations regarding the GW's are rather poor in
 physical significance.

\vskip0.70cm
\begin{center}
\noindent \emph{\textbf{APPENDIX}}
\end{center}

% \vskip0.10cm
\par \emph{$\alpha$}) I report here the summary of a communication by
I.Leonor at LIGO Scientific Collaboration meeting, March 23, 2005,
entitled ``Searching for GRB-GWB coincidence during LIGO science
runs''.
\par \emph{Summary}:
\begin{itemize}
\item[-] developed scheme for searching for GRB-GWB coincidence in
near real time \item[-] looking forward to S5 run with $\sim$ 100
GRB triggers in one year of coincident run \item[-]  performed
search for short-duration GW bursts coincident with S4, S3, and S2
GRB's using crosscorrelation method \item[-] sample probability
distribution consistent with null hypothesis. --
\end{itemize}

\par The LIGO scholars are technically very clever, but evidently
they cannot discover a non-existent object as a GW.

\par \emph{$\beta$}) On \emph{arXiv:gr-qc/0505029 v1} (6 May 2005) we
can read a paper of 23 pages, written by 395 LIGO-researchers all
over the world, entitled ``Upper limits on gravitational wave
bursts in LIGO's second science run -- LIGO-P040040-07-R''.

\par Here are some sentences from the ABSTRACT: ``We perform a search for
 gravitational wave bursts using data from the second science run of
 the LIGO detectors, using a method based on a wavelet time-frequency decomposition.
This search is sensitive to bursts of duration much less than a
second and with frequency content in the 100-1100 Hz range. It
features significant improvements in the instrument sensitivity
and in the analysis pipeline with respect to the burst search
previously reported by LIGO. $[\ldots]$. No gravitational wave
signals were detected in 9.98 days of analyzed data. $[\ldots]$''.

\par At p.11 we read: ``The WaveBurst analysis applied to the S2 data yielded 16
coincidence events (at zero-lag). The application of the
\emph{r}-statistic cut rejected 15 of them, leaving us with a
single event that passed all the analysis criteria.''. And at
p.13: ``The investigation revealed that the event occurred during
a period of strongly elevated acoustic noise at Hanford lasting
tens of seconds, as measured by microphones placed near the
interferometers. $[\ldots]$. The source of the acoustic noise
appears to have been an aircraft.'' --

\par In spite of the repeated failures, the LIGO scientists are
still hopeful. \emph{Spes ultima dea}. --

\par \emph{$\gamma$}) On \emph{arXiv:gr-qc/0505042 v1} (10 May 2005)
the above 395 scholars have published an article (7pp.) entitled
``Search for Gravitational Waves from Primordial Black Hole Binary
Coalescences in the Galactic Halo''. From the ABSTRACT: ``We use
data from the second science run of the LIGO gravitational-wave
detectors to search for the gravitational waves from primordial
black hole (PBH) binary coalescence with component masses in the
range 0.2--$1.0 M_\odot$. $[\ldots]$. No inspiral signals were
found.'' -- Obviously: both GW's and BH's are non-existent objects
\cite{9}. The so-called \emph{observed} BH's are enormously
massive bodies restricted in relatively small volumes -- as it can
be demonstrated by a careful scrutiny of the concerned papers
\cite{10}. --

\par \emph{$\delta$}) Again the mentioned 395 scientists on
\emph{arXiv:gr-qc/0505041 v1} (12 May 2005): ``Search for
gravitational waves from galactic and extra--galactic binary
neutron stars'' (20pp.). From the ABSTRACT: ``We use 373 hours
($\approx$ 15 days) of data from the second science run of the
LIGO gravitational-wave detectors to search for signals from
binary neutron star coalescences within a maximum distance of
about 1.5 Mpc, a volume of space which includes the Andromeda
Galaxy and other galaxies of the Local Group of galaxies.
$[\ldots]$. No inspiral gravitational wave events were identified
in our search.'' The conclusion of the paper is the following
(p.19): ``In this paper, we have presented a data analysis
strategy that could lead to a detection of gravitational waves
from binary neutron star inspirals. The methods used to validate
the search illustrate the subtleties of the analysis of several
detectors with different sensitivities and orientations. Moreover,
the experience gained by following up the largest coincident
triggers will be crucial input to investigations of event
candidates that are identified in future searches.'' An Italian
jest says: \emph{Chi vive sperando muore cantando}. --

\end{document}